\documentclass[12pt]{iopart}

\usepackage{amssymb}
\usepackage{bm}
\usepackage{epsfig}
\usepackage{iopams}
\renewcommand{\P}{\mathcal{P}}
\renewcommand{\e}{\mathrm{e}}                  %% eulersche Zahl
\newcommand{\ket}[1]{|#1\rangle}             %% ket-Vektor
\newcommand{\bra}[1]{\langle #1|}            %% bra-Vektor
\newcommand{\braket}[2]{\langle #1|#2\rangle}%% Skalarprodukt
\renewcommand{\d}[1]{\mathrm{d}#1}           %% Differential
\renewcommand{\Im}{\mathrm{Im}}              %% Imaginaerteil 
\newcommand{\mt}[1]{\mathrm{#1}}             %% fr Text-Indizes

\renewcommand{\bs}[1]{\boldsymbol{#1}}        %% bold greek letters
\newcommand{\wh}{\widehat}
\begin{document}

\title[Scattering of two-level atoms by delta lasers]{Scattering of two-level atoms by delta lasers: Exactly solvable models in atom optics}
\author{D Seidel$^1$, J G Muga$^1$ and G C Hegerfeldt$^2$}

\address{$^1$ Departamento de Qu\'{\i}mica-F\'{\i}sica, Universidad del
Pa\'{\i}s Vasco, Apartado Postal 644, 48080 Bilbao, Spain\\}
\address{$^2$ Institut f\"ur Theoretische Physik, Universit\"at G\"ottingen,
Friedrich-Hund-Platz 1, 37077 G\"ottingen, Germany}
\ead{\mailto{dirk\_seidel@ehu.es},\mailto{jg.muga@ehu.es}}

\begin{abstract}
We study the scattering of two-level atoms at narrow laser fields, modeled by a $\delta$-shape intensity profile. The unique properties of these potentials allow us to give simple analytic solutions for one or two field zones. Several applications  are studied: a single $\delta$-laser may serve as a detector model for atom detection and arrival-time measurements, either by means of fluorescence or variations in occupation probabilities. We show that, in principle, this ideal detector can measure the particle density, the quantum mechanical flux, arrival time distributions or local kinetic energy densities.
Moreover, two spatially separated $\delta$-lasers are used to investigate quantized-motion effects on Ramsey interferometry. 
\end{abstract}

\pacs{03.75.Be, 42.50.Vk, 03.65.Ta}

%\maketitle

\section{Introduction}

In standard scattering theory, $\delta$-potentials are useful models to study the properties of scattering solutions, bound states \cite{Morse-book, Gottfried-book} or inverse scattering \cite{ChaSa-book, Lamb-book}, and to check the validity or illustrate different approximations, concepts
or techniques \cite{M1,M2}. Because of the unique properties of these potentials, analytical solutions are easy to obtain without the use of special functions or extensive calculations. In many cases, $\delta$-potentials give a qualitative understanding to more complex scattering systems. For example, the Kronig-Penney model in solid state physics \cite{Berezin}, consisting of a lattice of $\delta$-functions, is very successful in describing energy gaps for the free electron gas in crystals. Further examples are finite lattices \cite{Joan},
time-dependent $\delta$-interactions \cite{propagator}, nonlinear delta interactions 
\cite{Molina}, or 
exactly solvable models of few-body and many-body systems \cite{Takahashi-book, La-AJP-1983, M3}. The Tonks-Girardeau gas of one-dimensional repulsive bosons subjected to effective $\delta$-interactions \cite{Girardeau}, in particular, 
has been recently realized experimentally with ultracold atoms,  
which are quite suitable for delta-interaction models because 
of their large de Broglie wavelengths.     
   
In all the above applications the particles are formally structureless. However, due to the rapid experimental progress in quantum optics and atom optics, in particular in the limit of ultracold conditions, there is a recent interest in solving multichannel scattering problems at localized fields taking into account the internal level structure of atoms. For example, a series of papers has been devoted to the reflection and transmission of slow atoms from micromaser barriers \cite{micromaser}. However, exact solutions for a nonresonant barrier soon become unwieldy, although two-channel recurrence relations may be used to put multiple barrier problems down to the single barrier case \cite{SeiMu-preprint-2006}.
For that reason, the study of $\delta$-laser models is extremely useful to understand atom-field interactions on a manageable level when taking into account the atomic quantized motion. A $\delta$-laser will not necessarily reproduce a real setup in its full glory but 
%besides its pedagogical convenience 
it provides a valuable tool to check approximations, concepts and ideal limits of operational quantities. A first analysis of the multi-channel $\delta$-potential problem can be found in \cite{FeVa-EJP-1996}. 

In this paper we derive first the stationary solutions for the scattering of two-level atoms by one or two $\delta$-laser fields, including a treatment of spontaneous decay. The results are then applied to atom detection, to quantum time measurements and to matter-wave interferometry.

\section{Interaction between a two-level atom and a $\delta$-laser}

\subsection{Stationary solutions}

\begin{figure}
\centering
\epsfxsize=9cm \epsfbox{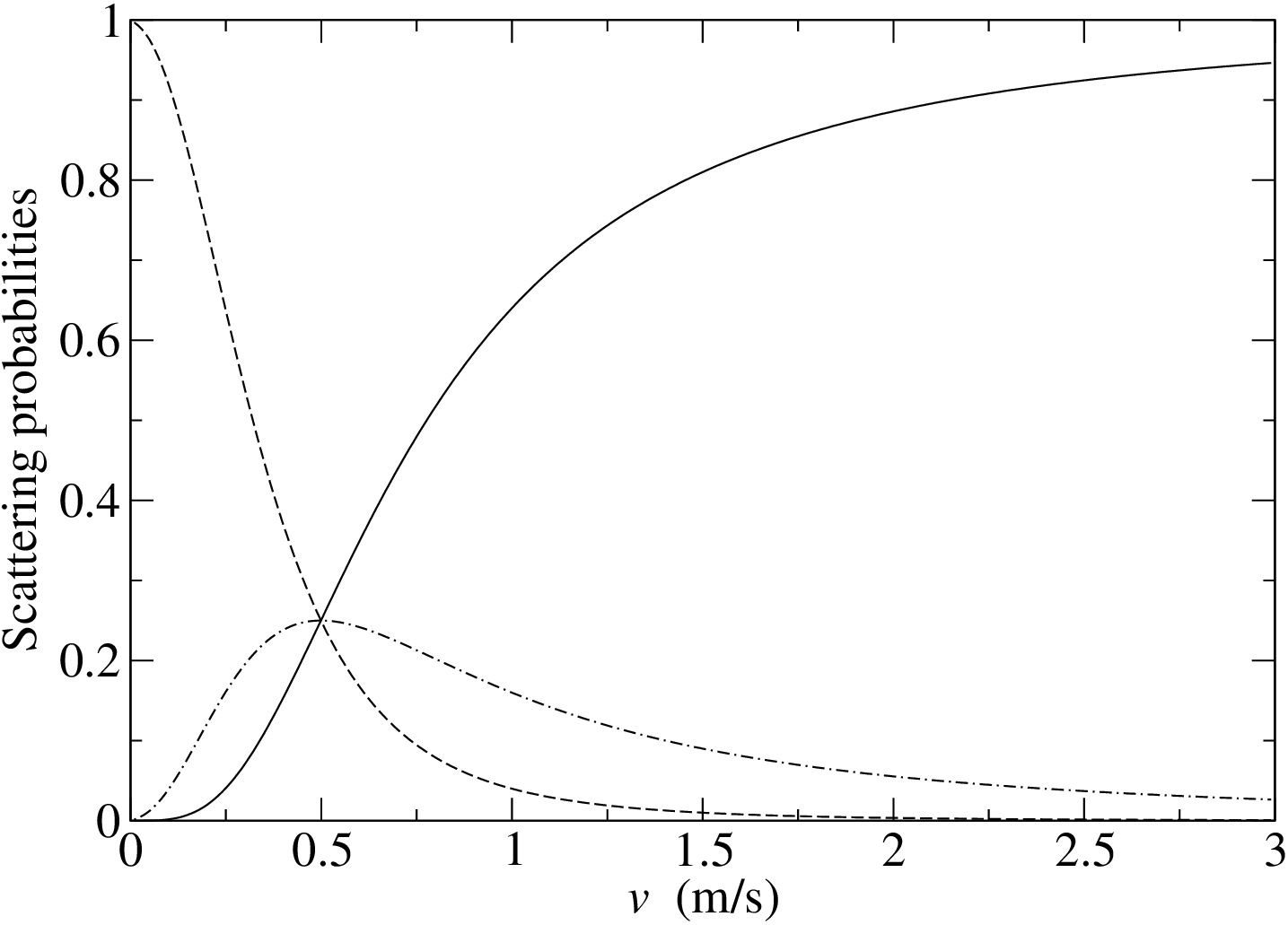}
\caption{Scattering probabilities of a two-level atom ($^{87}$Rb, $m = 1.4\times 10^{-25}\,\mt{kg}$) at a $\delta$-laser, $|t_{11}|^2$ (solid line), $|r_{11}|^2$ (dashed line), $\frac{q}{k}| r_{12}|^2 = \frac{q}{k}|t_{12}|^2$ (dashed-dotted line). Parameters are: $u = 0.1\,\mt{MHz} \times 10\,\mu\mt{m} = 1\,\mt{m/s}$, $\gamma=0$, $\Delta = 100\,\mt{Hz}$.}
\label{fig:deltascattering}
\end{figure}

First, let us consider a moving two-level atom with internal states $\ket{1} = {1 \choose 0}$ and $\ket{2} = {0 \choose 1}$, interacting with a narrow, non-resonant and $\delta$-like laser  field located at $x=\xi$. The direction of the laser beam is assumed to be perpendicular to the motion of the atom. 
Note that in our one-dimensional treatment only the longitudinal motion is quantized but transverse momentum transfer is neglected. These recoil effects have been studied in detail in the context of Ramsey interferometry \cite{Borde}. Their neglection is reasonable if the atoms are confined in a narrow waveguide in the Lamb-Dicke regime, moving freely only in one direction \cite{SeiMu-preprint-2006}.
The whole problem becomes time-independent by using the standard field-adapted interaction picture. To incorporate decay of the upper level $\ket{2}$ with a decay rate $\gamma$, we use the quantum jump approach \cite{QJA,DaEgHeMu-PRA-2002}. Here, the evolution of the atom before the first emission of a photon and in the dipole and rotating-wave approximation is given by the non-Hermitian (``conditional") Hamiltonian
\begin{equation} \label{eq:Hamiltonian}
H_\mt{c} = \frac{\wh{p}^{\, 2}}{2m} \bs{1}_2 + \frac{\hbar u}{2} \delta(\wh{x} - \xi)
\Biggl( \begin{array}{cc} 0&1\\1&0 \end{array} \Biggl) - \frac{\hbar}{2} 
\Biggl( \begin{array}{cc} 0&0\\0&\rmi\gamma+2\Delta \end{array}\Biggr),
\end{equation}
where $\Delta = \omega_\mt{L} - \omega_{21}$ is the detuning between laser and atomic frequency and $u$ controls the strength of the $\delta$-laser and has dimensions of velocity. An approximate physical realization corresponds to a square laser profile of width $l$ and Rabi frequency 
$\Omega$ such that $\Omega l=u$, with the $\delta$-limit achieved 
as $\l\to 0$. The eigenfunctions of $H_\mt{c}$ with energy $E = \hbar^2 k^2/2m =mv^2/2$ for a ground state plane wave incoming from the left are given by
\begin{equation}
\bPhi_k(x) = \cases{ {\rme^{\rmi kx} + r_{11} \rme^{-\rmi kx} \choose r_{12} \rme^{-\rmi qx}},\qquad x \leq \xi \\  {t_{11} \rme^{\rmi kx} \choose t_{12} \rme^{\rmi qx}}, \qquad x \geq \xi,}
\end{equation}
where $q^2 = k^2 + m(\rmi\gamma + 2\Delta)/\hbar$ and the scattering amplitudes $r_{ij}, t_{ij}$ are determined by the matching conditions at $x=\xi$. They read
\numparts
\begin{eqnarray} \label{eq:R11}
r_{11} &=& - m^2 u^2\,\exp(2\rmi k\xi)/d,\\ \label{eq:T11}
t_{11} &=& 4\hbar^2 k q/d, \\ \label{eq:R12}
r_{12} &=& -2\rmi \hbar m k u\, \exp[\rmi (k+q)\xi]/d, \\ \label{eq:T12}
t_{12} &=& - 2\rmi \hbar m k u\, \exp[\rmi (k-q)\xi]/d,
\end{eqnarray}
\endnumparts
where $d=4\hbar^2 k q + m^2 u^2$.
According to (\ref{eq:R12}) and (\ref{eq:T12}), excitation by scattering at the $\delta$-laser is suppressed for slow atoms ($v \ll u/2$) and for fast atoms ($v \gg u/2$), see figure~\ref{fig:deltascattering}, whereas maximal excitation is achieved for $v \approx u/2$ ($v = u/2$ for $\gamma = \Delta = 0$).

%For later purpose, we also give the scattering solution for an incoming excited state plane wave,
%\begin{equation}
%\bPhi_k(x) = \cases{ {R_{21} \rme^{-\rmi kx} \choose \rme^{\rmi q x} + R_{22} \rme^{-\rmi qx}},\qquad x \leq \xi \\  {T_{21} \rme^{\rmi kx} \choose T_{22} \rme^{\rmi qx}}, \qquad x \geq \xi,},
%\end{equation}
%where the matching conditions give $T_{22} = T_{11}$, $R_{22} = R_{11}$ and $R_{21} = T_{21} = -2i\hbar m qu/(4\hbar^2 kq + m^2 u^2)$.

\subsection{Particle detection by spontaneous emission}

The quantum description of detection and in particular of arrival times is still subject of discussion, see \cite{Muga-book} for reviews.
In a series of recent papers, the detection process has been modeled by means of the fluorescence from a two-level atom interacting with a spatially confined laser field \cite{DaEgHeMu-PRA-2002,NaEgMuHe-JPB-2003,DaEgHeMu-JPB-2003}. This model has led to new insight into the connection between operational quantities (defined for the system
in interaction with a measuring apparatus) 
and ideal arrival-time distributions (defined for the particle in isolation). It is shown in the following that $\delta$-lasers can be used to model point detectors and that the outcome of such a detector can be related to various {\it ideal} quantities by taking appropriate limits. Other point-like detector models with $\delta$-potentials have been studied in \cite{AoHoNa-PRA-2000, HeSeiMuNa-PRA-2004, Bo-PRA-2004}.

We consider a two-level atom in the ground state, incoming from the far left and impinging on a $\delta$-laser localized at $\xi = 0$. Before the first spontaneous emission, the atom is described by a conditionally evolved wave packet %\cite{DaEgHeMu-PRA-2002},
\begin{equation} \label{eq:Psi_t}
\braket{x}{\mathbf{\Psi}(t)} = \mathbf{\Psi}(x,t) = \int_0^\infty \d k\, \widetilde{\psi}(k)\,\e^{-\rmi
    \hbar k^2 t/2m} \mathbf{\Phi}_k(x).
\end{equation}
Here $\braket{k}{\psi} = \widetilde{\psi}(k)$ denotes the momentum amplitude that the initial ground state wave packet would have at $t=0$ in the absence of the laser.
Note that for $\gamma\neq 0$ the norm of $\ket{\mathbf{\Psi}(t)}$ decreases in time and it gives the probability to observe no photon until time $t$.
The interaction with the laser field will probably excite the atom and it may emit a photon subsequently. The probability density of the first photon emitted at time $t$ is given by \cite{DaEgHeMu-PRA-2002} 
\begin{equation} \label{eq:Pi(t)}
\Pi(t) = -\frac{\d}{\d t}\Vert \mathbf{\Psi}(t)\Vert ^2 =  \gamma \int_{-\infty}^\infty \d x |\braket{2}{\mathbf{\Psi}(x,t)}|^2.
\end{equation}
To relate this operational quantity to an ideal detection process, we consider limiting cases with respect to the parameters $\gamma$, $u$ and $\Delta$. Let us assume zero detuning, $\Delta = 0$, and strong coupling between the atom and the detector to assure immediate detection.  To correct for the decrease in the detection amplitude due to ground state transmission and reflection off the field we define the normalized detection rate
\begin{equation} \label{eq:normalization}
\Pi_\mt{N}(t) = \Pi(t) \left(\int_{-\infty}^\infty \d t\,\Pi(t)\right)^{-1}.
\end{equation}
%
%see \cite{DaEgHeMu-PRA-2002} for a discussion. 
More sophisticated normalization procedures on the level of operators can be defined to preserve the bilinear form properties of the distribution as we will show below. 

With respect to $\Pi_\mt{N}$  we obtain two main results: First, for strong decay, $\hbar\gamma \gg E$ and $\hbar\gamma \gg mu^2$, the normalized detection rate in leading order becomes
\begin{equation} \label{eq:density}
\Pi_\mt{N}(t) = \langle v^{-1} \rangle^{-1} |\psi_\mt{free}(0,t)|^2,
\end{equation}
where $\psi_\mt{free}(x,t) = \int \d k\,\widetilde{\psi}(k) \exp(\rmi k x -\rmi \hbar k^2 t/2m)$ is the freely evolving wavepacket and the normalization constant is the mean inverse velocity of this packet. 
%\footnote{For sufficiently monochromatic packets this is approximately the inverse of the mean velocity.}.
Thus, by means of the fluorescence measurement one obtains the density of the unperturbed wave packet at the position where the $\delta$-laser is located. With respect to arrival time measurements we note that (\ref{eq:density}) has been shown in \cite{MuSeiHe-JCP-2005} to be the zeroth order of an expansion of Kijowski's arrival-time distribution \cite{Kijowski} for nearly monochromatic wave packets and it has been recently used to study arrival times in the presence of interactions \cite{AoHoNa-PRA-2000}.

The second result is obtained in parameter regimes for which $mu^2 \gg \hbar\gamma \gg E$. In that case the leading order yields
\begin{equation} \label{eq:KED}
\Pi_\mt{N}(t) = \frac{2}{p_0} \langle \wh{K} (0) \rangle_t,
\end{equation}
where $\wh{K}(\xi) = \wh{p} \delta(\wh{x}-\xi) \wh{p}/2m$ is the local kinetic energy operator at $x=\xi$, the expectation value is taken over the freely evolving packet $\psi_\mt{free}(t)$ and $p_0$ is its mean momentum. Thus, a strongly coupled $\delta$-laser provides a way to measure the local kinetic energy density of the particle which is an important quantity in chemical physics. A recent discussions can be found in \cite{MuSeiHe-JCP-2005,AyPaNa-IJQC-2002}. Note that in (\ref{eq:density}) and (\ref{eq:KED}) the particle density and the local kinetic energy density at an arbitrary point $x=\xi$ are obtained by simply shifting the laser field.

\subsection{Operator normalization}

The normalization procedure given in (\ref{eq:normalization}) does not provide a  distribution which is bilinear in the state. This is in contrast to 
ideal bilinear quantities such as the quantum mechanical flux or Kijowski's distribution \cite{Kijowski}. To preserve these bilinear form properties, Brunetti and Fredenhagen proposed a normalization formalism on the level of operators \cite{BruFre-PRA-2002}. Their idea has been recently applied to quantum arrival times \cite{HeSeiMu-PRA-2003}. In this section we will briefly review the basic idea and we will present results for the first-photon distribution at $\delta$-lasers.

To begin, we note that (\ref{eq:Pi(t)}) can be written as an expectation value of an operator $\widehat{\Pi}_t$ over the unperturbed initial ground state at time $t=0$. Indeed, if $U_\mt{c}(t,t_0) = \exp(\rmi H_0 t/\hbar)\exp[-\rmi H_\mt{c}(t-t_0)/\hbar] \exp(-\rmi H_0 t_0/\hbar)$ is the conditional time development operator corresponding to $H_\mt{c}$ in an interaction picture with respect to $H_0 = \wh{p}^{\,2}/2m$ then one finds
\begin{equation}
 \Pi(t) = \bra{1}\bra{\psi} \widehat{\Pi}_t \ket{\psi}\ket{1},
\end{equation} 
where the operator $\widehat{\Pi}_t$ is given by
\begin{equation}
\wh{\Pi}_t = -\frac{\d}{\d t} U_\mt{c}(t,-\infty)^\dagger U_\mt{c}(t,-\infty).
\end{equation}
In this notation, the conditionally developed state $\ket{\mathbf{\Psi}(t)}$ given in (\ref{eq:Psi_t}) reads $\ket{\mathbf{\Psi}(t)} = \exp(-\rmi H_0 t/\hbar) U_\mt{c}(t,-\infty) \ket{\psi}\ket{1}$. In analogy to (\ref{eq:normalization}) one considers the operator-valued normalization
\begin{equation} \label{eq:Bdef}
 \int_{-\infty}^\infty \d t\,\wh{\Pi}_t = \wh{1} - U_\mt{c}(\infty,-\infty)^\dagger U_\mt{c}(\infty,-\infty) =: \wh{B},
\end{equation}
where $\wh{B}$ is defined on the set of incoming states with internal ground state. In the language of scattering theory, $\wh{B} = \wh{1} - \wh{S}^\dagger \wh{S}$, where $\wh{S}$ is the usual scattering operator connecting incoming and outgoing asymptotic states. Note that $\wh{B}$ averaged over the incoming asymptotic state provides the total detection probability.
Following the derivation of \cite{HeSeiMu-PRA-2003}, we obtain for our problem of a $\delta$-laser interaction the following representation of $\wh{B}$ in $k$-space:
\begin{equation}
 \bra{1}\bra{k} \wh{B} \ket{k'}\ket{1} = [1 - r_{11}^*(k)r_{11}(k') - t_{11}^*(k) t_{11}(k')] \delta(k-k').
\end{equation}
Now, on the incoming states, one can define the positive distribution
\begin{equation} \label{eq:PiON1}
 \Pi^\mt{ON}_1(t) = \bra{1}\bra{\psi} \wh{B}^{-1/2} \wh{\Pi}_t \wh{B}^{-1/2} \ket{\psi}\ket{1}.
\end{equation}
Since $|r_{11}(k)|^2 + |t_{11}(k)|^2 < 1$, the inverse square-root of $\wh{B}$ exists. From (\ref{eq:Bdef}) it is obvious that $\Pi_1^\mt{ON}(t)$ is normalized. This allows us to consider the strong coupling regime $E \ll \hbar \gamma, mu^2$, since operator normalization compensates for the reflection and transmission losses caused by ground state scattering. We obtain in leading order the ideal arrival-time distribution of Kijowski at $x=0$ \cite{Kijowski},
\begin{eqnarray}
\Pi_1^\mt{ON}(t) \simeq \Pi_\mt{K}(0;t) = \frac{\hbar}{2\pi m} \Biggl| \int \d k\, \widetilde{\psi}(k) \sqrt{k} \exp(-\rmi \hbar k^2 t/2m) \Biggr|^2,\nonumber \\ \qquad\qquad\qquad\qquad\qquad\qquad E \ll \hbar \gamma, mu^2\quad\mt{and}\quad \hbar\gamma \sim mu^2.
\end{eqnarray}
Operationally, the operator normalization can be viewed in the following way: the acting of the operator $\widehat{B}^{-1/2}$ in (\ref{eq:PiON1}) can be understood as a filtering of the incident state before the actual detection process takes place. This filtering amplifies (in a relative way) the very slow and very fast momentum components which are preferentially reflected or transmitted without emitting a photon, see \cite{HeSeiMu-PRA-2003} for a discussion. 

Instead of (\ref{eq:PiON1}) one may also choose the not manifestly positive Rivier symmetrization rule
\begin{equation}
 \Pi^\mt{ON}_2(t) = \frac{1}{2}\bra{1}\bra{\psi} \wh{B}^{-1} \wh{\Pi}_t + \wh{\Pi}_t \wh{B}^{-1} \ket{\psi}\ket{1}.
\end{equation}
We note that it is much more difficult to obtain an operational understanding of this expression. 
%, thus we consider it here as a more fundamental alternative to (\ref{eq:PiON1}).
In the same regime as above, $\Pi^\mt{ON}_2(t)$ yields in leading order the quantum mechanical flux at $x=0$,
\begin{equation}
 \Pi_2^\mt{ON}(t) \simeq J(0;t), \qquad E \ll \hbar \gamma, mu^2\quad\mt{and}\quad \hbar\gamma \sim mu^2,
\end{equation}
which is the classical result for an arrival-time expression but its quantum version is not a positive distribution even for particles with only positive momenta because of the backflow effect \cite{BraMe-JPA-1994}. A more reasonable operational approach to the quantum mechanical flux by means of fluorescence has been given in \cite{DaEgHeMu-PRA-2002}.

\subsection{Particle detection by variations in occupation probabilities}

In this section we present results for another operational quantity that may be used for particle detection, namely variations in the occupation probabilities. For a two-level atom, let us consider
\begin{equation}
\Pi(t) = \frac{\d P_2}{\d t} = \frac{\d}{\d t} \Vert \braket{2}{\mathbf{\Psi}(t)} \Vert^2,
\end{equation}
which is measurable by probing the excited state at different times. For vanishing decay, $\gamma=0$, the time evolution is unitary and with the Hamiltonian (\ref{eq:Hamiltonian}) $\Pi(t)$ becomes
\begin{equation}
\Pi(t) = u\, \Im \Bigl(\braket{1}{\mathbf{\Psi}(x=0,t)} \braket{\mathbf{\Psi}(x=0,t)}{2} \Bigr)
\end{equation}
Interestingly, the normalized version of this quantity leads to the same ideal distributions as in the fluorescence measurement, but now for different parameter regimes. For $mu^2 \ll E \ll \hbar\Delta$ we find the density,
\begin{equation}
\Pi_\mt{N}(t) = \langle v^{-1} \rangle^{-1}|\psi_\mt{free}(0,t)|^2,
\end{equation}
whereas for $E \ll mu^2$, $E\ll\hbar\Delta$ and $mu^2/(\hbar\Delta) \sim 1$ we end up with the local kinetic energy density,
\begin{equation}
\Pi_\mt{N}(t) = \frac{2}{p_0} \langle \wh{K} (0) \rangle_t.
\end{equation}
As above, both results are with respect to the freely evolving wave packet.

\section{Interaction between a two-level atom and two $\delta$-lasers}

\subsection{Stationary solutions}

Let us now consider a two-level atom, interacting with two separated laser fields with $\delta$-shape. This setup provides a toy model to study Ramsey's method of matter-wave interferometry with separated fields including effects of quantized motion. 
The conditional Hamiltonian for the atom is given in analogy to (\ref{eq:Hamiltonian}) by
\begin{equation} \label{eq:Hamiltonian2}
H_\mt{c} = \frac{\wh{p}^{\, 2}}{2m} \bs{1}_2 + \frac{\hbar u}{2} \bigl[\delta(\wh{x}) + \delta(\wh{x}-L)\bigr]
\Biggl( \begin{array}{cc} 0&1\\1&0 \end{array} \Biggl) - \frac{\hbar}{2} 
\Biggl( \begin{array}{cc} 0&0\\0&\rmi\gamma+2\Delta \end{array}\Biggr),
\end{equation}
where $\xi =0$ has been set for convenience and $L$ is the distance between both lasers.
The asymptotic eigenfunctions of $H_\mt{c}$ for a plane wave incoming from the left and in the ground state are given by
\begin{equation}
\bPhi_k(x) = \cases{ {\rme^{\rmi kx} + R_{11} \rme^{-\rmi kx} \choose R_{12} \rme^{-\rmi qx}},\qquad x \leq 0 
%\\ {A_+ \e^{ikx} + A_- \e^{-ikx} \choose B_+ \e^{iqx} + B_- \e^{-iqx}},\qquad 0 \leq x \leq L 
\\  {T_{11} \rme^{\rmi kx} \choose T_{12} \rme^{\rmi qx}}, \qquad x \geq L,}
\end{equation}
where now the amplitudes have to be determined by the matching conditions at $x=0$ and $x=L$ and they read
\numparts
\begin{eqnarray} \label{eq:2delta1}
R_{11} &=& -m^2 u^2\bigl[4\hbar^2 kq (1+\e^{2\rmi kL} + 2\e^{\rmi(k+q)L}) \nonumber\\
&&\qquad\qquad\qquad\qquad + m^2 u^2(\e^{2\rmi kL}-1)(\e^{2\rmi qL} -1) \bigr]/D,\\ \label{eq:2delta2}
R_{12} &=& -2\rmi\hbar k mu \left[4\hbar^2 kq(1+\e^{\rmi(k+q)L}) + m^2 u^2(\e^{2\rmi kL}-1)(\e^{2\rmi qL} -1) \right]/D,\nonumber\\&& \\ \label{eq:2delta3}
T_{11} &=& 4\hbar^2 kq \e^{-\rmi kL} \left[4\hbar^2 kq \e^{\rmi kL} + 2\rmi m^2 u^2 \e^{\rmi qL}\sin(kL)\right]/D,\\ \label{eq:2delta4} 
T_{12} &=& -8\rmi\hbar^3 k^2 q mu \e^{-\rmi qL}(\e^{\rmi kL} + \e^{\rmi qL})/D
\end{eqnarray}
\endnumparts
with the common denominator
\begin{equation} \label{eq:2delta5}
D = 16 \hbar^4 k^2 q^2 +8\hbar^2 kq m^2 u^2(1+\e^{\rmi(k+q)L}) + m^4 u^4 (\e^{2\rmi kL}-1)(\e^{2\rmi qL} -1).
\end{equation}
From (\ref{eq:2delta1})--(\ref{eq:2delta5}) it can be seen that $T_{12}$ and $R_{12}$ tend to zero for $k\to 0$ and for $k\to\infty$. Thus, also for a double-laser setup the excitation probability vanishes for very slow and for very fast particles.

\subsection{Ramsey interferometry}

Atom interferometry based on Ramsey's method with separated fields \cite{Ramsey} is an important tool of modern precision measurements and the basis of atomic clocks. 
An essential feature of the observed Ramsey interference fringes is that its width is simply the inverse of the time taken by the atoms to cross the intermediate region.
This implicates the desire for very slow (ultracold) atoms \cite{Salomon}. But, if the kinetic energy becomes comparable with the atom-field interaction energy, one has to take into account the quantized center-of-mass motion of the atom and the well-known semiclassical results have to be corrected \cite{SeiMu-preprint-2006}. A useful toy model to study analytically these effects is provided by the double $\delta$-laser setup.

The measured quantity in a Ramsey interferometry experiment is the transmission probability of excited atoms, $P_{12}(\Delta)$, as a function of the detuning $\Delta$. This function is easily obtained in the semiclassical regime where $E \gg mu^2$, $E \gg \hbar \Delta$ and the center-of-mass motion can be treated independently of the internal dynamics. In this regime, $P_{12}(\Delta)$ has been derived in \cite{Ramsey} for rectangular field shapes. Applying the $\delta$-limit to this expression yields
\begin{equation} \label{eq:P12scl}
 P_{12}^\mt{scl}(\Delta) = 4 \sin^2\left(\frac{u}{2v} \right) \cos^2\left(\frac{u}{2v} \right) \cos^2\left(\frac{\Delta L}{2 v} \right).
\end{equation} 
%
%where $k_u = mu/\hbar$. 
From (\ref{eq:P12scl}) we find the Ramsey fringes to be of $\cos^2$-shape and their width to be $2\pi/T$, where $T=L/v$ is the semiclassical crossing time of the intermediate region. 
%Moreover, full excitation for resonant interaction is obtained if $k_u/k = \pi/2$ which is the translation of the usual $\pi/2$-condition to the $\delta$-laser case. 
Note that in contrast to rectangular field shapes $P_{12}^\mt{scl}$ does not drop to zero for large detuning.

\begin{figure}
\centering
\epsfysize=5cm \epsfbox{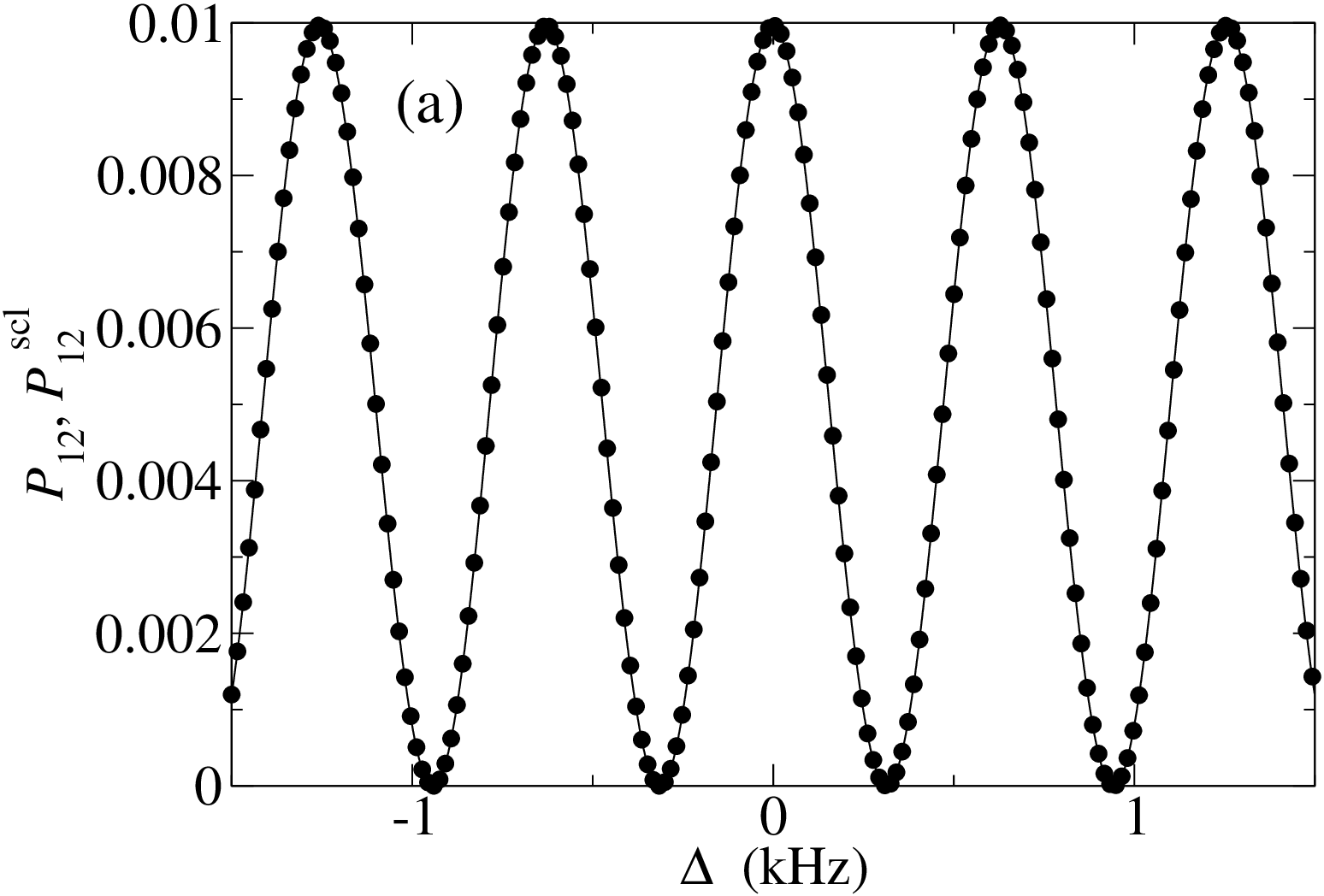}\hfill  \epsfysize=5cm \epsfbox{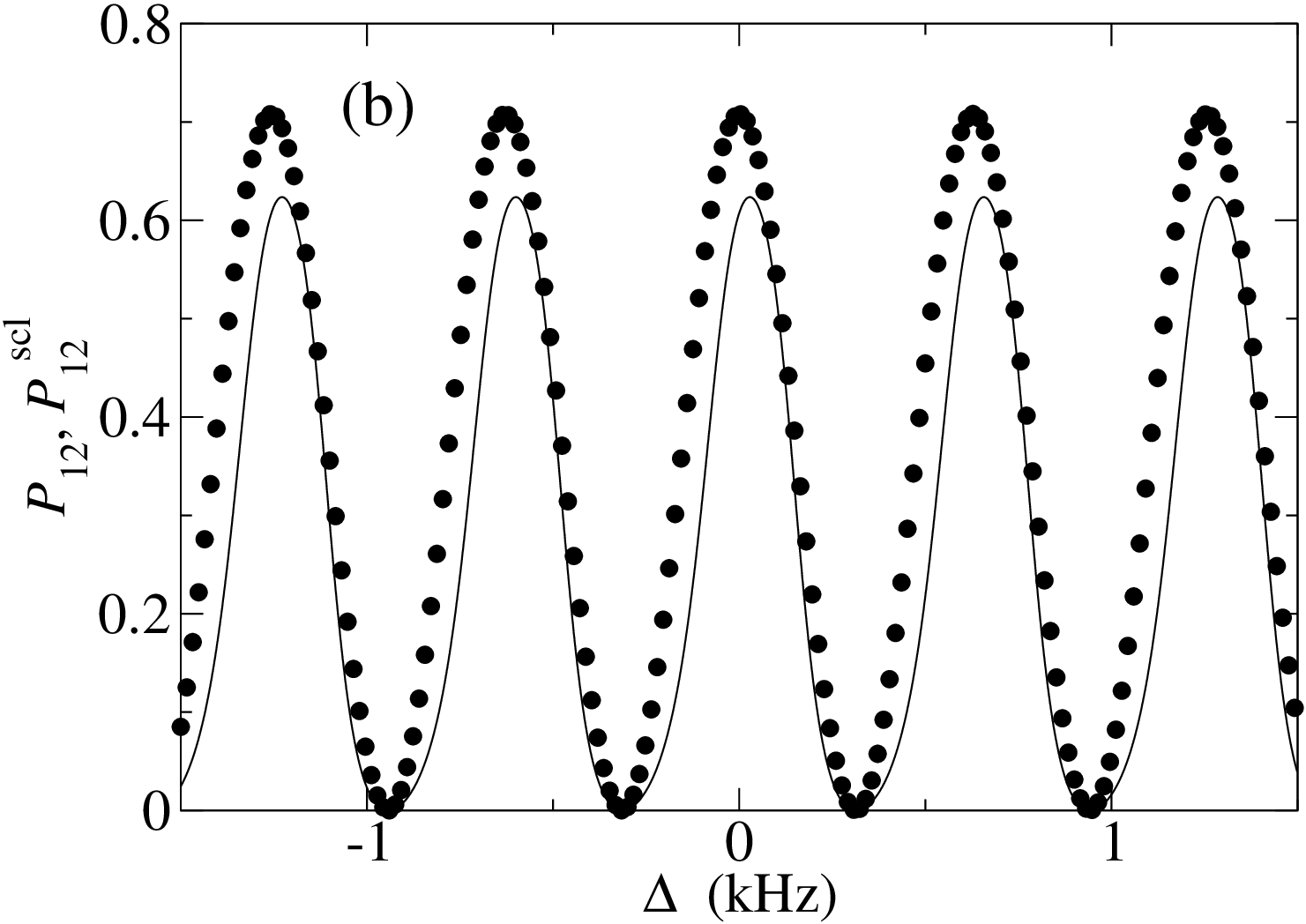}
\epsfysize=5cm \epsfbox{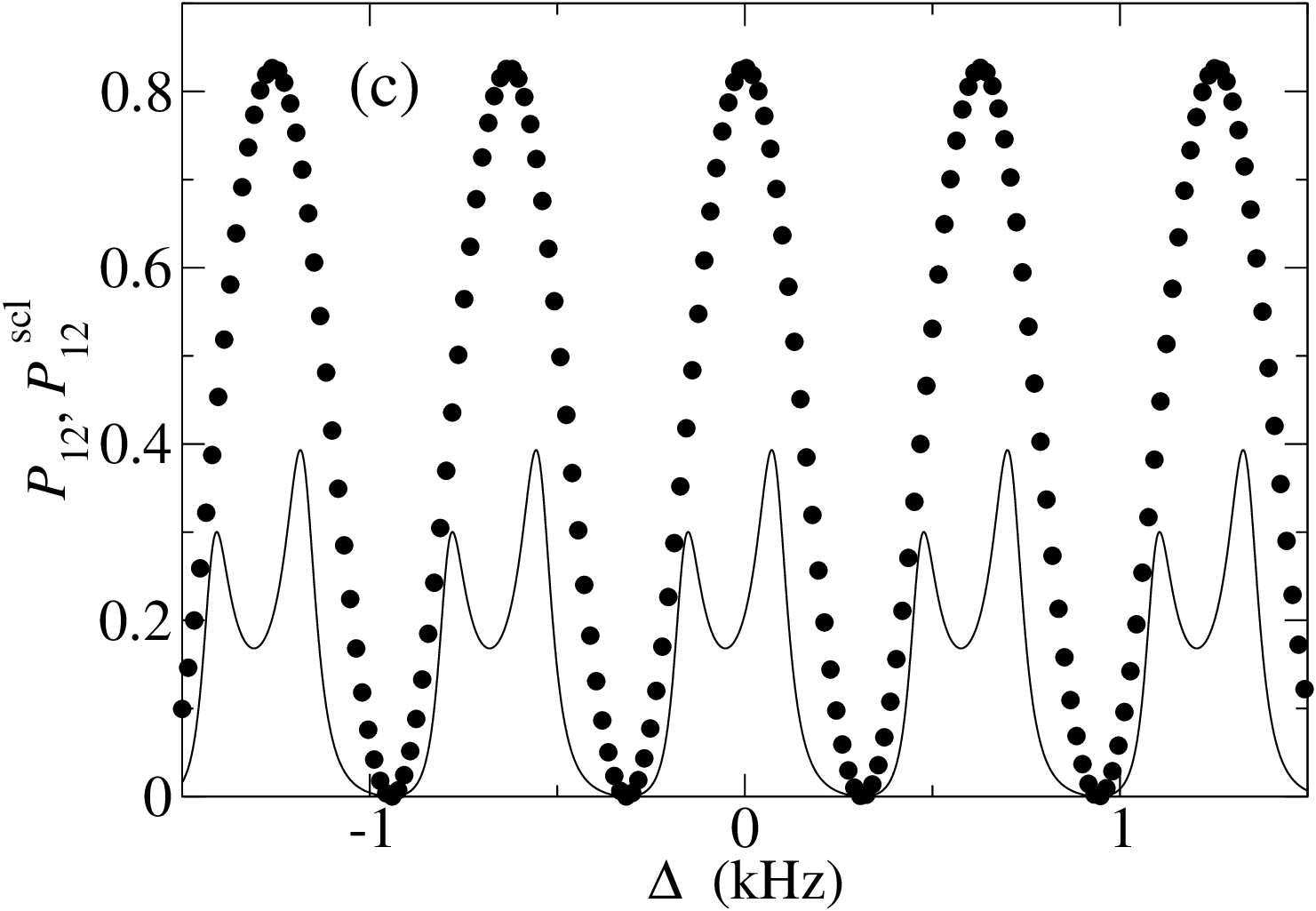}\hfill  \epsfysize=5cm \epsfbox{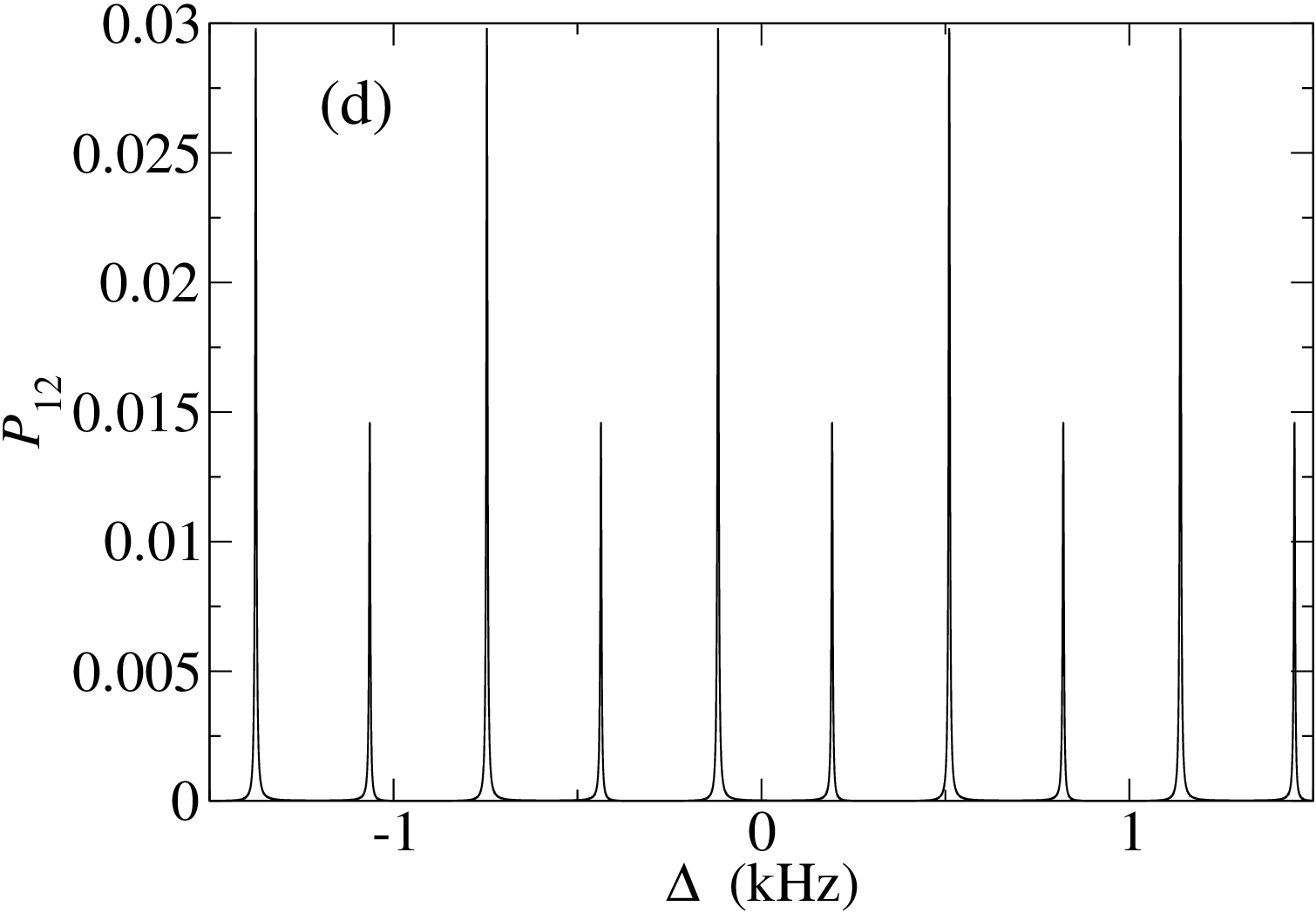}
\caption{Ramsey interference fringes for a $^{87}$Rb atom passing through two $\delta$-laser fields: Exact quantum result (solid line, (\ref{eq:P12})) and semiclassical approximation (circles, (\ref{eq:P12scl})). Parameters are $\gamma=0$, $u = 1\,\mt{m/s}$, $L = 0.01\,\mt{s} \times v$ and (a) $v = 10\,\mt{m/s}$, (b) $v = 1\,\mt{m/s}$, (c) $v = 0.5\,\mt{m/s}$, (d) $v = 0.1\,\mt{m/s}$.}
\label{fig:P12}
\end{figure}

However, if the kinetic energy of the atom is comparable with the interaction energy, the semiclassical approach is not valid anymore and a full quantum mechanical solution is required. In that case the excitation probability is given in terms of the scattering amplitudes derived above. Since the outgoing excited state component becomes evanescent for $\Delta$ smaller then the critical value $\Delta_\mt{cr} = -\hbar k^2/2m$ due to the wavenumber $q$, we finally obtain the quantum result to be
\begin{equation} \label{eq:P12}
P_{12}(\Delta) = \frac{q}{k} |T_{12}|^2 \qquad\mt{for}\qquad  \Delta > \Delta_\mt{cr}
\end{equation}
and zero elsewhere \cite{SeiMu-preprint-2006}. From this one can study the transition from the semiclassical to the quantum regime. For $k\to\infty$, (\ref{eq:P12}) reduces to $P_{12} = u^2 v^{-2} \cos^2[\Delta L/(2v)]$ which agrees in leading order of $u/v$ with the semiclassical expression (\ref{eq:P12scl}), as expected.
A numerical example is shown in figure~\ref{fig:P12}a. For slower and slower atoms, $P_{12}$ differs from $P_{12}^\mt{scl}$ and changes its behavior from interference fringes to scattering resonances, see figures~\ref{fig:P12}b-d. The reason for this is the increasing dominance of quantum reflections at the field that have been neglected in the semiclassical description.
A more detailed study of this transition has been given in \cite{SeiMu-preprint-2006}. It is important to note that for kinetic energies of the order of the interaction energies the maxima of the quantum fringes are slightly displaced with respect to the semiclassical result which may become important as a systematic frequency uncertainty in future time standards using ultracold atoms.

\section{Discussion}

We have investigated analytically the one-dimensional scattering of two-level atoms at one and two $\delta$-laser fields. The relatively simple expressions for the scattering amplitudes make this a valuable toy model to study applications of atom optics in the ultracold regime where the kinetic energy is comparable with the atom-field interaction energy. 

The main objective of our analysis is to provide more insight into to concepts and ideal limits of models in atom optics.
A physical realization of the $\delta$-laser, of course, might be difficult since this requires the de Broglie wavelength of the atom to be larger than the laser beam width and $\hbar\Omega$ to be larger than the kinetic energy. However, with subrecoil cooling techniques and translational temperatures of the order of nK the corresponding de Broglie wavelength is of the order of $\mu$m which is not far away from possible laser beam waists.

We have shown that a single $\delta$-laser may serve as a particle detector either through fluorescence measurements or through variations in the atomic occupation probabilities. The outcome of such a detector has been related to several ideal quantities, such as the probability density, the flux, Kijowski's arrival time distribution or local kinetic energy densities, by considering different parameter regimes and normalization procedures. These relations are of fundamental importance, since the antithetic questions ``How to measure ideal quantities?" and ``What kind of quantities an experiment really measures?" are only insufficiently answered for quantum detection processes and quantum time measurements so far.

Moreover, by means of the double $\delta$-laser setup we have studied Ramsey interferometry including quantum reflections at the fields. We obtained an exact quantum mechanical solution for the interference fringes that agrees in the weak coupling regime with the well-known expression of Ramsey but shows deviations for stronger coupling or slower atoms. A peak shift of the central fringe at $\Delta = 0$ occurs which should be considered as a systematic frequency uncertainty in future atomic clocks with ultraslow atoms. For the extreme case that the atomic kinetic energy is much smaller than the atom-field interaction energy, the standard fringes are completely suppressed and the excitation probability $P_{12}(\Delta)$ exhibits a Fabry-Perot resonance structure.

\section*{Acknowledgments}

This work has been supported by Ministerio de Edu\-ca\-ci\'on y Ciencia (BFM2003-01003) and UPV-EHU (00039.310-15968/2004). D.S.\ acknowledges a fellowship within the Postdoc-Programme of the German Academic Exchange Service (DAAD).

\end{document}